# Log-Normal Distribution of Single Molecule Fluorescence Bursts in Micro/Nano-Fluidic Channels

Lazar L. Kish [(1)], Jun Kameoka [(1)], Claes G. Granqvist [(2)], Laszlo B. Kish [(1)]

[(1)] *Department of Electrical and Computer Engineering, Texas A&M University, College Station, TX 77843-3128, USA*

[(2)] *Department of Engineering Sciences, The Ångström Laboratory, Uppsala University, P. O. Box 534, SE-75121 Uppsala, Sweden*

The width and shape of photon burst histograms pose significant limitations to the identification of single molecules in micro/nano-fluidic channels, and the nature of these histograms is not fully understood. To reach a deeper understanding, we performed computer simulations based on a Gaussian beam intensity profile with various fluidic channel diameters and assuming (*i*) a deterministic (noise-free) case, (*ii*) photon emission/absorption noise, and (*iii)* photon noise with diffusion. Photon noise in narrow channels yields a Gaussian burst distribution while additional strong diffusion produces skewed histograms. We use the fluctuating residence time picture [Phys. Rev. Lett. **80**, 2386-2388 (1998)] and conclude that the skewness of the photon number distribution is caused by the longitudinal diffusive component of the motion of the molecules as they traverse the laser beam. In the case of strong diffusion in narrow channels, this effect leads to a log-normal distribution. We show that the same effect can transform the separate peaks of the photon burst histograms of multiple molecule mixtures into a single log-normal shape.

Single fluorescent color based single-molecule detection is of great interest in a variety of fields[1-4] including air quality monitoring.[5] Micro- and nanofluidic devices are important, and one of the novel platforms and methodologies—denoted MAPS (microfluidic system for analyzing proteins in a single complex)—was presented recently by Chou *et al.*[1] MAPS relies on quantum dots that are selectively attached to proteins which are drifting in a micro/nano-fluidic channel through a laser beam wherein they are excited and emit a group of photons (referred to as a "photon burst"). Different types or numbers of quantum dots attached to various proteins result in corresponding photon burst sizes, and thus such a system can identify single protein molecules. In the present Letter, we address the size distribution of the photon burst and, for the sake of simplicity, mainly discuss the case of MAPS. However our results are generally valid for any single molecule fluorescence experiment with micro/ nanofluidic flow as carrier.

Ideally, a photon burst should be a fixed number corresponding to the given protein molecule and the attached quantum dots. However, the physical situation is less clear-cut as a consequence of random fluctuations ensuing from photon noise (shot noise due to absorption and emission) and diffusion (random walk) of molecules. Repeated experiments yield a histogram with a non-zero width of the photon number distribution,[1] and similar results can be seen in other types of single fluorescent color based single-molecule experiments with micro/nanofluidics.[2,3] Some of these distributions are well understood;[3] for example the Poisson statistics of photon absorption and emission (photonic noise) results in a Gaussian distribution of the number of photons in the burst. One should note that even in a hypothetical case without photonic noise there would be a distribution with non-zero width due to the Gaussian beam profile; this distribution is caused by the random initial location of the molecule within the fluidic channel. The situation results in different burst numbers as a consequence of passing the beam at different distances from its center where its intensity is at a maximum. The effect yields a distribution with a peak close to zero photon number.[3]

However, recent experiments[1] have shown that the distribution is often skewed and has a long tail extending toward large photon numbers with resemblance to a log-normal distribution.[6-8] Such distributions occur when the logarithm of a random variable has a Gaussian distribution and have been associated with nanoparticles and nanotechnology for some 35 years.[6] The log-normal density function is given by[9]

$$f(x) = \frac{1}{\sqrt{2\pi}\,\ln\sigma} \exp\left[-\frac{\left(\ln x - \ln\mu\right)^2}{2(\ln\sigma)^2}\right], \tag{1}$$

where $\mu$ is the geometric mean and $\sigma$ is a dimensionless standard deviation. A log-normal distribution is usually confirmed via a logarithmic probability plot wherein the cumulative distribution of the log-normal variable gives a straight line. Below we study the role of diffusion and present simulations to explain the origin of the log-normal like shape of the burst distribution. Specifically, we show that the log-normal distribution is the result of residence time fluctuations due to diffusion. A similar situation was identified for gas evaporation of nanoparticles;[6-8] however the latter case did not have shot noise in the particle growth and hence a new study is needed to investigate the more complex conditions in the current experiments involving micro/nano-fluidics.

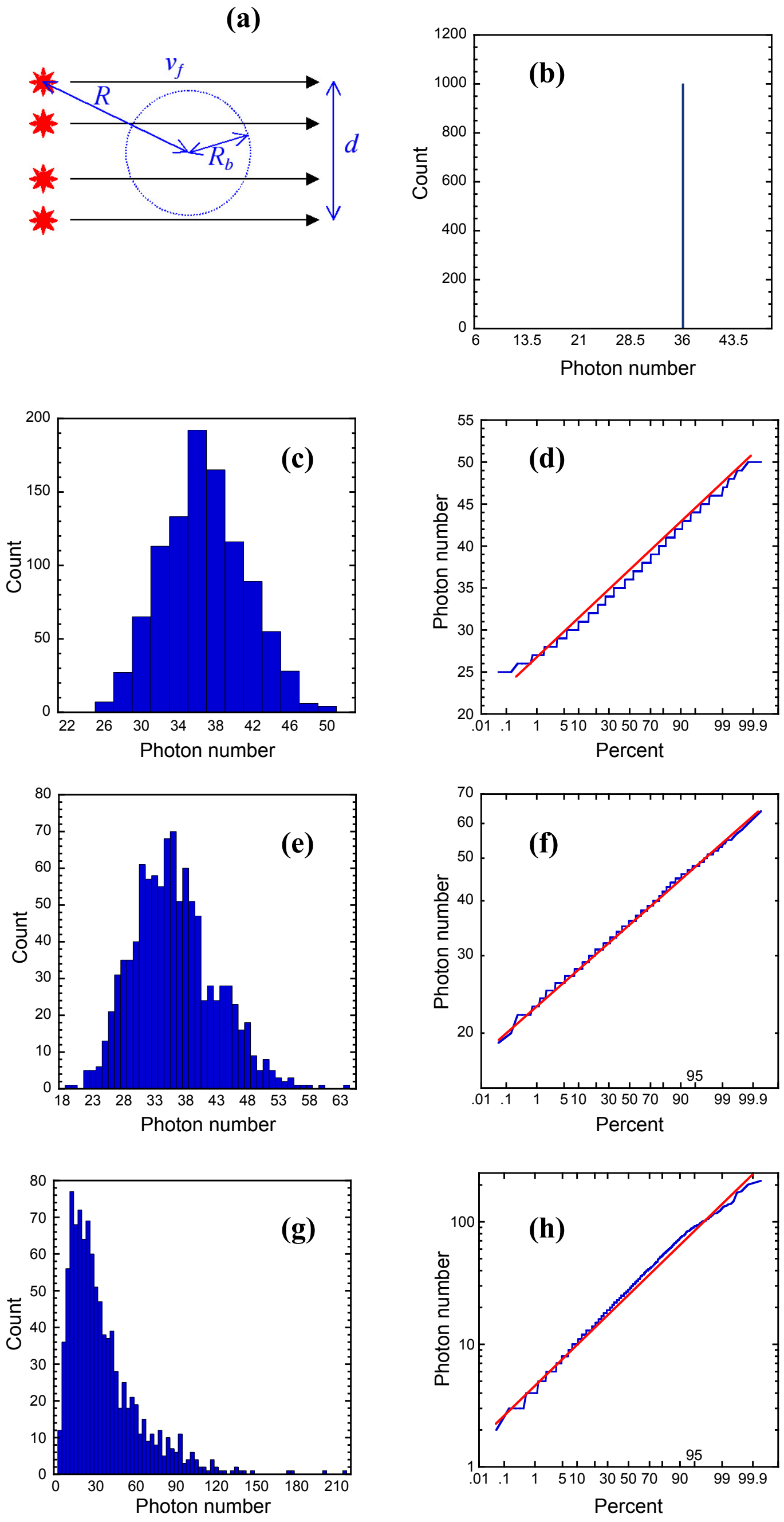

**Figure 1.** Simulations for the case of a narrow fluidic channel with $R_b/d$ = 16. **(a)** Illustration of the geometry of the experimental situation with a Gaussian beam. **(b)** Histogram of the randomness-free case (no photon noise and no diffusion). **(c)** Photonic noise without diffusion ($D = 0$) and **(d)** normal probability plot of its cumulative distribution. **(e)** Photonic noise with weak diffusion ($D = 0.1$) and **(f)** its log-normal probability plot. **(g)** Photonic noise with strong diffusion ($D = 5.63$) and **(h)** its log-normal probability plot.

We assume that the molecules proceed with a constant drift velocity due to laminar flow and by diffusion (*i.e.*, by random walk) superimposed on that motion. The excitation and emission follow Poisson statistics, where the rate of excitation is proportional to the local intensity of laser light and the rate of emission by the excited molecules is constant (*i.e.*, induced emission is neglected).

The excitation probability $P_{ex}$ of the molecule follows a Rayleigh distribution with Poisson statistics, *i.e.*,

$$P_{ex} = P_0 \exp\left(-\frac{R^2}{2R_b^2}\right) , \qquad (2)$$

where $P_0$ (<< 1) is the excitation probability at the center of the beam during one time step, $R$ is the instantaneous distance of the molecule from the center of the beam, and $R_b = 320$ is the effective radius of the Gaussian beam; *cf*. Fig. 1(a). Without diffusion, the molecules follow the laminar flow as in Fig. 1(a), where a constant velocity ($v_f$) flow-profile is supposed. As computer simulations use arbitrary units, a normalization of the numbers is needed to interpret their physical meanings. With the chosen flow velocity $v_f = 1$ in the diffusion-free case, it takes 640 time steps for the molecule to drift through the effective diameter $2R_b = 640$ of the laser beam.

The strength $D$ of the diffusion is quantified by the square of the ratio of the effective displacement of the diffusive motion (random walk) along the channel and the effective diameter $2R_b = 640$ of the laser beam during the 640 time steps, while supposing zero flow. One should note that $D$ is proportional to the diffusion coefficient of the experiment.

Figure 1(b) shows simulation results for a narrow fluidic channel ($R_b/d$ = 16) under the hypothetical case of no random fluctuations, *i.e.*, no photon noise (deterministic absorption and emission) and no diffusion ($D$ = 0). This hypothetical situation results in an ideal single-line histogram.

Simulation results for photonic noise with a *narrow* fluidic channel ($R_b/d$ = 16) and *no diffusion* ($D = 0$) are shown in Figs. 1(c,d). The histogram of the burst size distribution is given in Fig. 1(c). The normal probability plot of the cumulative distribution in Fig. 1(d), which is a rigorous check of Gaussian statistics, shows a normal (Gaussian) distribution at least within the range of 1 to 99 %; the straight line indicates an exact Gaussian distribution.

Figures 1(e,f) illustrate simulation results for photonic noise with a *narrow* fluidic channel ($R_b/d$ = 16) and *weak diffusion* ($D$ = 0.1). The histogram of the burst size distribution in Fig. 1(e) indicates that the original Gaussian peak of Fig. 1(c) now exhibits some asymmetry. The logarithmic probability plot of the cumulative distribution in Fig. 1(f) signifies that this weakly skewed distribution is actually a good log-normal at least within the range 0.1 to 99.9%.

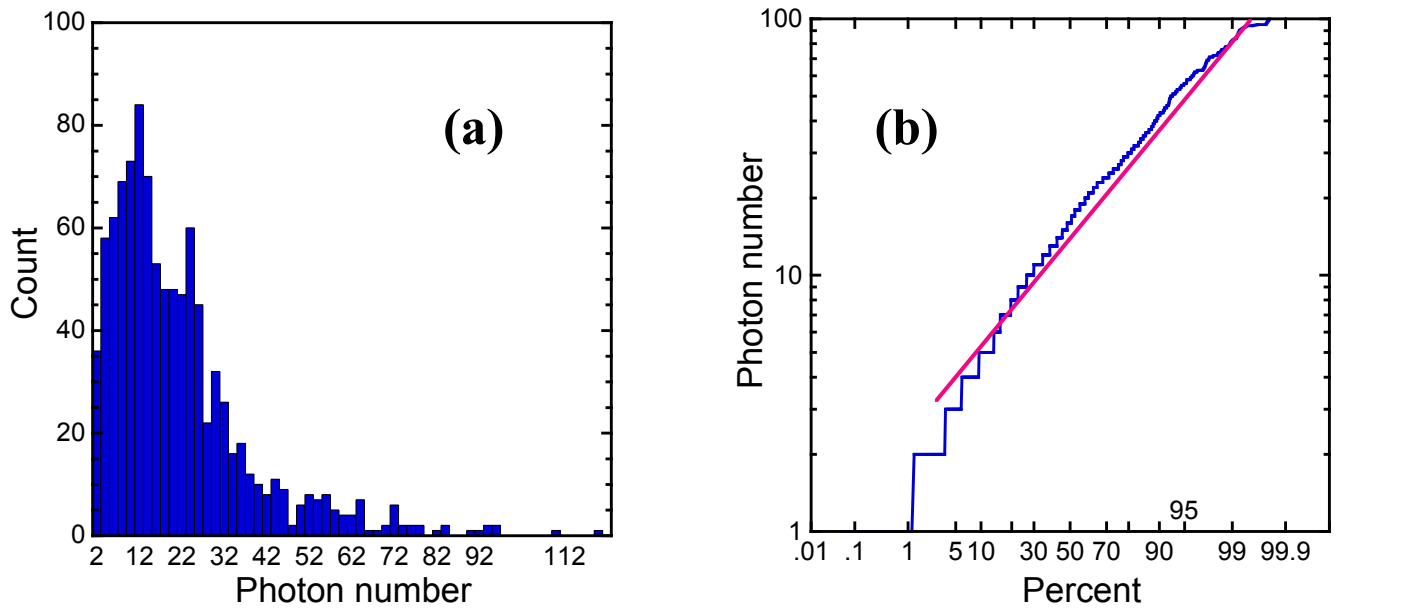

**Figure 2.** Simulations for the case of a wide fluidic channel ($R_b/d$ = 0.32) and strong diffusion ($D$ = 5.63). (**a**) Histogram of photonic noise and (**b**) its log-normal probability plot.

Simulation results for photonic noise with a *narrow* fluidic channel ($R_b/d$ = 16) and *strong diffusion* ($D$ = 5.63) are depicted in Figs. 1(g,h). The histogram of burst size distribution in Fig. 1(g) displays a strongly skewed distribution with no resemblance to the Gaussian of photon noise obtained without diffusion. The logarithmic probability plot of the cumulative distribution in Fig. 1(h) indicates a fairly log-normal distribution at least within the range 1 to 99%.

Corresponding results for photonic noise with a *wide* fluidic channel ($R_b/d$ = 0.32) and *strong diffusion* ($D$ = 5.63) are shown in Fig. 2. The histogram of burst size distribution in Fig. 2(a) shows a wide and skewed distribution. The logarithmic probability plot of the cumulative distribution in Fig. 2(b) signifies a roughly log-normal distribution only within the range 10 to 99%. Thus wide fluidic channels destroy part of the log-normality and lead to a wider burst size distribution.

Finally, Fig. 3 shows simulation results for three different molecule species where—due to the low concentration of molecules—only a single molecule moves in the laser beam at a given time. The histogram of the burst size distribution without diffusion ($D$ = 0) shows three maxima—centered around 100, 200 and 300 photons—in Fig. 3(a). The original histogram is very significantly modifies already at weak diffusion ($D$ = 0.13), as apparent from Fig. 3(b), and the second and third peaks are largely merged. As found from Fig. 3(c), the original histogram is totally destroyed at strong diffusion ($D$ = 1.8) and all of the three peaks are then combined. The logarithmic probability plot of the cumulative distribution in Fig. 3(d) indicates a log-normal distribution within the range of about 2 to 95%.

In conclusion, we have simulated the role of diffusion in single molecule micro/nano-fluidics experiments[3-6] and demonstrated that the skewness and the width of the distribution of photon numbers can be explained by residence time fluctuations in the excitation zone analogously to the case of nanoparticle growth with log-normal size distribution in gas evaporation.[7,8] These fluctuations are caused by the diffusive (random walk) component of the motion that is superimposed upon the drift provided by the flow of the fluid and lead to log-normal like size distributions under a wide range of experimental conditions for which diffusion plays a decisive role.

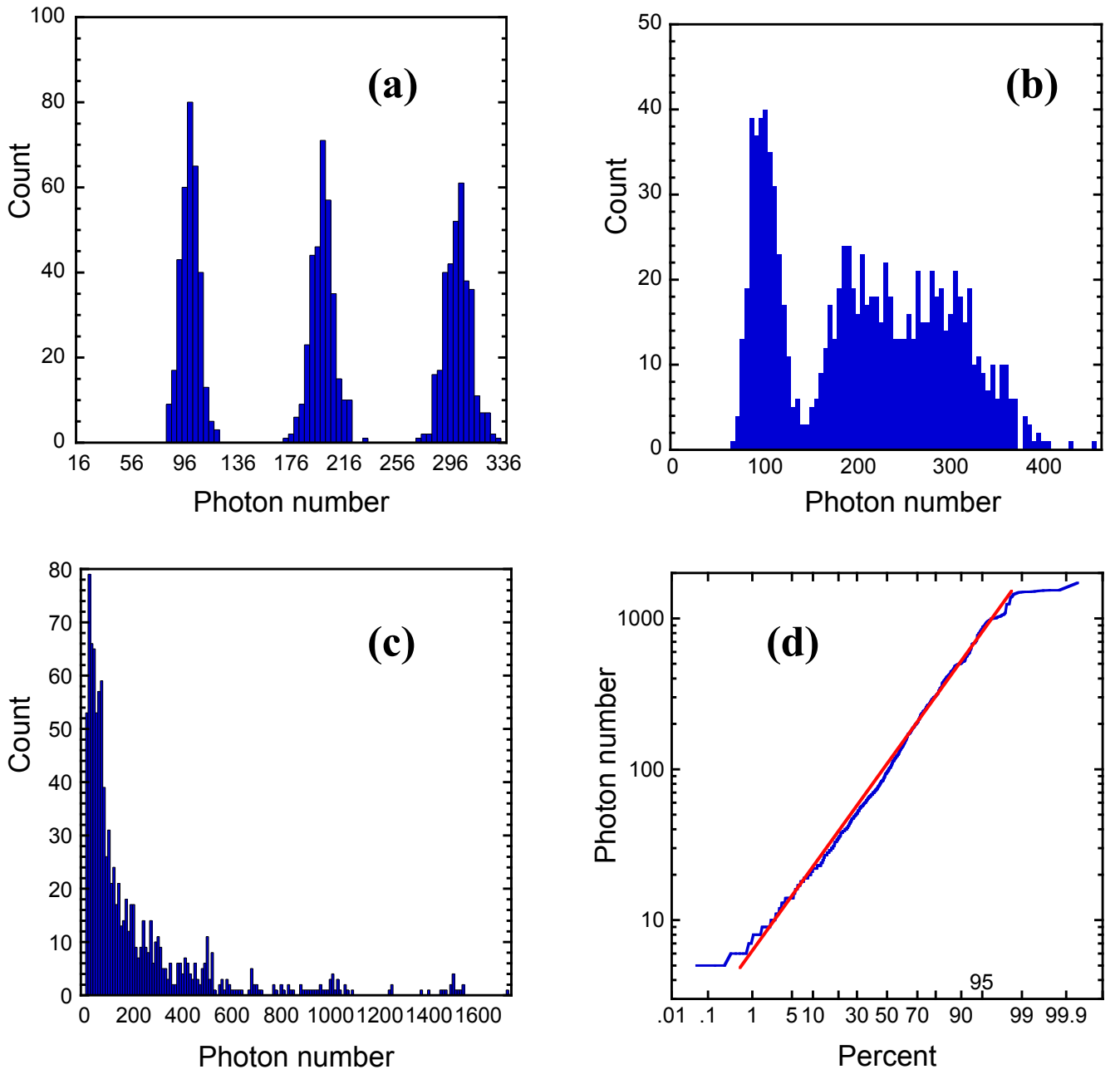

**Figure 3.** Simulations for three different molecule species as discussed in the main text. (**a**), **(b)** and **(c)** Histograms of burst size distributions at no diffusion ($D$ = 0), weak diffusion ($D$ = 0.013) and strong diffusion ($D$ = 1.6), respectively. (**d**) Log-normal probability plot for the data in (c).

Financial support was received from the European Research Council under the European Community's Seventh Framework Program (FP7/2007-2013)/ERC Grant Agreement 267234 ("GRINDOOR").